\documentclass[12pt,aps,tightenlines,showpacs,nofootinbib,letterpaper,superscriptaddress]{revtex4}
\usepackage{epsfig}
\usepackage{graphicx}
\usepackage{amsmath}
\usepackage{amsfonts}
\usepackage{amssymb}
\usepackage{bm}

\newcommand{\gammab}{\mbox{\boldmath$\gamma$}}
\newcommand{\nablab}{\mbox{\boldmath$\nabla$}}

\begin{document}
\title{Mass Varying Neutrinos, Quintessence, and the Accelerating Expansion of the Universe}
\author{Gennady Y. Chitov}
\affiliation{Department of Physics,
Laurentian University, Ramsey Lake Road, Sudbury, ON, P3E 2C6, Canada }
\author{Tyler August}
\affiliation{Department of Physics, Laurentian University,
Ramsey Lake Road, Sudbury, ON, P3E 2C6,
Canada }
\author{Aravind Natarajan}
\affiliation{McWilliams Center for Cosmology and Department of Physics,
Carnegie Mellon University, 5000 Forbes Ave, Pittsburgh, PA 15213, USA}
\author{Tina Kahniashvili}
\affiliation{McWilliams Center for Cosmology and Department of
Physics, Carnegie Mellon University, 5000 Forbes Ave, Pittsburgh,
PA 15213, USA}
\affiliation{Department of Physics, Laurentian
University, Ramsey Lake Road, Sudbury, ON, P3E 2C6, Canada }
\affiliation{National Abastumani Astrophysical Observatory, Ilia
Chavchavadze State University, 2A Kazbegi Ave, Tbilisi, GE-0160,
Georgia}
\date{\today}

\begin{abstract}
We analyze the Mass Varying Neutrino (MaVaN) scenario. We consider a minimal
model of massless Dirac fermions coupled to a scalar field, mainly in the
framework of finite temperature quantum field theory. We demonstrate that the
mass equation we find has non-trivial solutions only for special classes of
potentials, and only within certain temperature intervals. We give most of our
results for the Ratra-Peebles Dark Energy (DE) potential. The thermal
(temporal) evolution of the model is analyzed. Following the time arrow, the
stable, metastable and unstable phases are predicted. The model predicts that
the present Universe is below its critical temperature and accelerates. At the
critical point the Universe undergoes a first-order phase transition from the
(meta)stable oscillatory regime to the unstable rolling regime of the DE field.
This conclusion agrees with the original idea of quintessence as a force making
the Universe roll towards its true vacuum with zero $\Lambda$-term. The present
MaVaN scenario is free from the coincidence problem, since both the DE density
and the neutrino mass are determined by the scale $M$ of the potential.
Choosing $M \sim 10^{-3}~$eV to match the present DE density, we can obtain the
present neutrino mass in the range $m \sim 10^{-2}-1~$eV and consistent
estimates for other parameters of the Universe.
\end{abstract}

\maketitle
%
%
\section{Introduction}\label{Intro}
%
%

Neutrino mass related questions are of great interest for
particle physics as well as for cosmology (for reviews see Ref.
\cite{Dolgov} and references therein). Current upper limits on the
sum of neutrino masses from cosmological observations are of the
order of 1 eV
\cite{Lesgourgues:2006nd,Strumia:2006db,Hannestad:2006zg}, while
neutrino oscillations give a lower bound of roughly 0.01 eV
\cite{Eidelman:2004wy,Abele:2008zz}, making neutrino mass an
established element of particle physics. Furthermore,
understanding the origin of neutrino mass opens a window into
understanding physical processes beyond the standard model of
particle physics
\cite{Mohapatra07,Avignone:2007fu,Weinberg08,Dodelson}.

It is now well established that about seventy four percent of the Universe is
comprised of dark energy (DE) (for reviews see Ref. \cite{Peebles:2002gy} and
citation therein). The present stage of evolution of the Universe is governed
by this dominant DE contribution, and the Universe experiences an accelerating
expansion \cite{Carroll:2003qq,Copeland:2006wr}. The nature of DE is still
unknown, and it is one of the major questions of modern cosmology. There are,
broadly speaking, three major possibilities proposed to explain the DE
\cite{Peebles:2002gy}. Most straightforwardly, and in good agreement with the
current observational data, it can be present just as the cosmological constant
\cite{Peebles:2002gy}. Secondly, the DE can be accommodated in some framework
of the modified non-Einsteinian gravity theories (see, e.g.,
Refs.~\cite{Nojiri:2003ft,Carroll:2004de}). And lastly, following the original
proposals \cite{Ratra:1988,Wetterich:1988} on the DE originating from a scalar
field action similar to the inflaton field, there has been a lot of activity
in constructing and analyzing various trial scalar field Lagrangians to
model the DE \cite{Copeland:2006wr}. Note, that it is even unclear what kind of
scalar field potential governs the inflationary expansion of the Universe
\cite{Linde}, and as the result, the effective quantum field that adequately
describes inflation is still under debate \cite{Weinberg:2008hq}. A similar
observation can be drawn from analyzing many potentials proposed for the DE
action \cite{Copeland:2006wr}.

On the other hand, several cosmological and astrophysical observations imply
that about twenty two percent of the Universe consists of dark matter (DM)
\cite{Peebles:2002gy}, if we admit the general relativity theory of gravity.
Most probably DM is formed through massive weakly interacting particles
(WIMPs), and the nature of these particles is also still unknown. There are
several recent observations performed by PAMELA \cite{Mocchiutti:2009sj} and
GLAST missions which indicate DM particle annihilations \cite{Baltz:2008wd}.
Recently it was proposed that both these observations could be used to test
baryogenesis \cite{Kohri:2009yn} which is one of the important problems
of the standard particle physics model.

Another puzzling question in modern cosmology is the coincidence problem - the
density of DE is comparable to the present energy density of DM. In turn,
the latter is comparable (within the order of magnitude), to the
energy density of cosmological neutrinos \cite{Dolgov,Lesgourgues:2006nd}). Is
there a mechanism explaining this coincidence? A very convincing answer to this
question is given by the mechanism of DM mass generation via various types of
DM-DE couplings, ranging from Yukawa to more exotic ones.
\cite{Anderson:1997un,Hoffman:2003ru,Farrar:2003uw,Kawasaki:1991gn,GarciaBellido:1992de,Comelli:2003cv}
The mass of the DM particle in this approach is naturally time-dependent, and
they were coined Varying Mass Particles (VAMPs). Various DE--DM interaction
models have been constrained by observations of Supernovae type Ia \cite{SuNo},
the age of the Universe \cite{Franca:2003zg,Franca:2009xp,Huey:2004qv}, Cosmic
Microwave Background (CMB) anisotropies \cite{Wang:2006qw,Mainini:2007nq}, and
Large Scale Structure (LSS) formation \cite{LSS}.

Fardon, Nelson and Weiner elaborated on the VAMP mechanism in the context of
neutrinos. \cite{Fardon:2003eh}\footnote{The DE-neutrino coupling and the
baryogenesis constraints have been also studied also in Ref. \cite{Gu:2003er}}.
In their model the relic neutrinos, i.e., fermionic field(s), interact with a
scalar field via the Yukawa coupling. If the decoupled neutrino field is
initially massless, then the coupling generates a (varying) mass of neutrinos
in this DE-neutrinos model. This mass varying neutrino (MaVaN) scenario is
quite compelling, since it connects the origin of neutrino mass to the DE, and
solves the additional coincidence problem of why the neutrino mass and DE are
of comparable scales \cite{Peccei:2004sz}. (For more on the coincidence, see,
e.g. \cite{SolaEtalCoincidence}). To consider neutrinos as particles which get
their mass through the coupling is attractive for particle physics, as well as
for its cosmological consequences. However there are significant issues that
have to be resolved for the sake of viability of the MaVaN scenario. Most
notably, it has been shown \cite{Afshordi:2005ym} that the model of Ref.
\cite{Fardon:2003eh} suffers from a strong instability due to the negative
sound speed squared of the DE-neutrino fluid (see also
\cite{Kaplinghat:2006jk}).

Any DM-DE coupling induces observable changes in large scale structure
formation \cite{Pettorino:2008ez}. The main reason for this is due to the presence of
additional DM contributions (perturbations) in the equation of motion which
determines the dynamics of the scalar field. The changes in the dynamics are
drastic when massive neutrinos are coupled to DE \cite{Afshordi:2005ym}. In
this case the squared sound speed of the DE-neutrino fluid defined as $c_s^2 =
{\delta P}/{\delta \rho}$, (where $\delta$ represents the variation, and $P$
and $\rho$ are pressure and energy density of the DE-neutrino fluid) is
negative. The negative squared sound speed results in an exponential growth of
scalar perturbations. \cite{Bjaelde:2007ki,Bean,Valiviita:2008iv,Wetterich}

After the critique in Ref. \cite{Afshordi:2005ym}, the issue of
stability of the DE-neutrinos fluid  has been addressed by many
authors
\cite{Kaplinghat:2006jk,Wetterich,Brookfield,Bernardini,Takahashi:2005kw,Takahashi:2006jt,Spitzer:2006hm,
Fardon:2005wc,Ichiki:2007ng}. Various physical assumptions were
made in those references in order to avoid the exponential
clustering of neutrinos. In particular, to achieve stability,
proposals were put forward to make the DE-DM model more
complicated, e.g., by extending it to a multi-component scalar
field, or by promoting its supersymmetry.
\cite{Takahashi:2005kw,Fardon:2005wc} We however are not inclined
to pursue this line of thought and will explore the simplest
possible ``minimal'' model. As we will demonstrate, the
occurrence of the instability in the coupled DE-neutrinos model is
meaningful, and we will explore the physical implications of this
phenomenon. Note that Wetterich and co-workers \cite{Wetterich} have
already analyzed various implications of the instability in the
MaVaN model on the dynamics of neutrino clustering.

In this paper we re-address the analysis of the DE-neutrinos coupled model.
What is really new in our results, to the best of our knowledge, apart from a
consistent equation for the equilibrium condition, is the analysis of the
thermal (i.e. temporal) evolution of the MaVaN model and prediction of its
stable, metastable and unstable phases. The analysis of the dynamics in the
unstable phase results in, for the first time in the framework of the MaVaN
scenario,  a picture of the present-time Universe totally consistent with
observations. Our findings are in line with the original proposal
\cite{Ratra:1988,Wetterich:1988} of the DE potential (quintessence) to model the Universe slowly rolling towards its true vacuum
($\Lambda=0$). As it turns out, the present Universe, seen as a system of the
coupled DE (quintessence) field and fermions (neutrinos) is below its critical
temperature. It is similar to a supercooled liquid which has not crystallized
yet: its high temperature (meta)stable phase became unstable, but the new
low-temperature stable phase ($\Lambda=0$) is still to be reached. The
Afshordi-Zaldarriaga-Kohri instability corresponding to $c_S^2 <0$ is just
telling us this.

The rest of the paper is organized as follows: In Section~\ref{M&F} we give the
outlook of the model and formalism applied and derive the basic equations for
the coupled model. In Section~\ref{MasEq} we present the qualitative analysis
of the equation which yields the fermionic (neutrino) mass. Section~\ref{PRPot}
contains analysis of the coupled model with the Ratra-Peebles DE potential at
equilibrium. The dynamics of the model applied to the whole Universe is studied
in Section~\ref{Dynamics}. The results are summarized in the concluding
Section~\ref{Concl}.
%
%
%
\section{Model and Formalism. Basic Equations}\label{M&F}
%
%
%
%
\subsection{Outlook}\label{Outlook}
%
%
In this paper we focus on the case when the scalar field potential $U(\varphi)$
does not have a non-trivial minimum, and the generation of the fermion mass is
due to the breaking of chiral symmetry in the Dirac sector of the Lagrangian. A
non-trivial solution of the fermionic mass equation is a result of the
interplay between the scalar and fermionic contributions. We consider the most
natural and intuitively plausible Yukawa coupling between the Dirac and the
scalar fields.

The key assumption is that the fermionic mass generation can be obtained from
minimization of the thermodynamic potential. That is, the coupled system of the
scalar bosonic and fermionic fields is at equilibrium, at least at some
temperatures. This will be analyzed below more specifically. We assume the
cosmological evolution, governed by the scale factor $a(t)$ to be slow enough
that the coupled system is at equilibrium at a given temperature $T(a)$. Then the
methods of thermal quantum field theory \cite{Kapusta,Zinn-Justin} can be
applied.

This problem is rather well studied with quantum field theory and statistical
physics in different contexts \cite{Kapusta,Zinn-Justin,Greiner}. The major
conceptual difficulty in applying  quantum field-theoretical methods for the
dark-energy scalar field is the lack of  ``well-behaved" potentials
interesting for cosmological applications. For instance, a class of the
very popular inverse power law slow-rolling quintessence potentials
\cite{Copeland:2006wr} are singular at the origin. Consequently, the field
theory should be understood as a sort of effective theory, and we plan to
address this issue more deeply in our future work.

As far as the fermionic sector of the theory is concerned,
one needs to distinguish two different cases pertinent for neutrino applications:\\
(i) an equal number of fermions and antifermions, i.e., zero
chemical potential $\mu=0$; \\
(ii) a surplus of particles over antiparticles, and small non-zero chemical
potential.

For the bounds on the neutrino chemical potential, see Refs.
\cite{Dolgov,ChemPot}. If experiments confirm neutrinoless
double beta decay, i.e., that neutrinos are Majorana fermions,
then the lepton number is not conserved \cite{Avignone:2007fu},
and one cannot introduce a (non-zero) chemical potential. Then
case (i) above is applicable, proviso that the Majorana
fields are utilized instead of the Dirac ones. For the case (i)
with Dirac fermions the ground state corresponds to a complete
annihilation of  fermion-antifermion pairs, i.e. the fermions
completely vanish in the zero-temperature limit.

Assumption of the fermion-antifermion asymmetry and (conserving) particle
surplus, i.e., of a non-zero chemical potential, results in the fermionic
contributions which survive the zero-temperature limit. However the smallness
of the zero-temperature contribution renders this issue rather academic.
Indeed, for the neutrinos we are interested in this study, by assuming the
\textit{maximal} particle surplus $n_\circ \sim 115~ {\rm cm}^{-3}$, one gets the
Fermi momentum $k_F \sim 3 \cdot 10^{-4}~$eV. For $m \sim 10^{-2}~$eV, one
obtains $\mu(T=0) =\varepsilon_F = \sqrt{k_F^2+m^2}= m +
\mathcal{O}(10^{-4}~\mathrm{eV})$. This results in a non-trivial vacuum with
the particle surplus frozen within an extremely narrow Fermi shell $m \leq
\varepsilon \leq \varepsilon_F$. Thus, trying to grasp the essential physics in
this study from possibly the simplest ``minimal model", we assume the fermions
to be described by a Dirac spinor field with zero chemical potential.

In this work we will use the standard methods of general
relativity and finite-temperature quantum field theory
extended for fields living in a spatially flat Universe with the
Friedmann-Lema\^itre-Robertson--Walker (FLRW) metric where the
line element is $ds^2=dt^2-a^2(t)d \mathbf{x}^2$. Here $t$ is the
physical time and $a(t)$ is the scale factor, which can be
obtained from the Friedmann equations \cite{Weinberg08,Dodelson}
\begin{eqnarray}
H^2 (t)&=& \left(\frac{\dot a}{a}\right)^2 = \frac{8 \pi G}{3} \rho_{\rm{tot}}\
, \label{friedmann1}
\\
{\dot H}(t) + H^2(t)=\frac{\ddot a}{a} &=& -\frac{4\pi G}{3} (\rho_{\rm{tot}} +
3P_{\rm{tot}})\ . \label{friedmann2}
\end{eqnarray}
Eqs. (\ref{friedmann1})-(\ref{friedmann2}) also lead to the continuity equation
\begin{equation}
\label{Frid0}
   \dot{\rho}_{\rm tot}+ \frac{3 \dot{a}}{a}(\rho_{\rm tot}+P_{\rm tot})=0~.
\end{equation}
Here the dot represents the physical time derivative and $\rho_{\rm tot} $ and
$P_{\rm tot}$ are the total energy density and pressure of the Universe. In
accordance with the (standard) $\Lambda$CDM model,  the Universe is assumed to
consist of (1) DE, (2) cold DM (CDM) made of weakly interacting massive
particles, presumably  $M_{\rm DM}>1 \sim 10~$GeV, (3) photons, and (4)
baryons. The DM and baryon density parameters today are $\Omega_{\rm DM} =
\rho_{\rm DM}(t_{\rm now})/\rho_{\rm cr} \approx 0.22$ and $\Omega_b =
\rho_b(t_{\rm now})/\rho_{\rm cr} \approx 0.04$. Here $\rho_{\rm cr} =
3H_0^2/(8\pi G) = 8.1 h^2 \times 10^{-47}~{\rm GeV}^4$ is the critical density
today, $t_{\rm now}$ defines the current time, $H_0 =2.1 h \times 10^{-42}~{\rm
GeV}$ is the present Hubble parameter, $G$ is the Newton constant, and $h
\approx 0.72$ is the Hubble parameter in units of 100 km$/$sec$/$Mpc. The photon
contribution to the energy density today can be neglected. The flatness of the
Universe leads to the relative energy density of the DE-neutrino coupled fluid
$\Omega_{\varphi \nu} \approx  0.74$. To ensure the accelerated expansion of
the Universe today, the r.h.s. of Eq.~(\ref{friedmann2}) must be positive at
$t=t_{\rm now}$.

In this paper we will not assume the existence of the cosmological constant
$\Lambda$, as the $\Lambda$CDM model suggests. Instead we accept the hypothesis
of the dynamical dark energy described by a scalar field. This is a bold
assumption and a highly debatable issue. We vindicate our approach \textit{a
posteriori} by the consistent picture we arrive at the end. For a review and/or
alternative approaches, see, e.g.,  Refs.
\cite{Copeland:2006wr,SolaEtal1,SolaEtal2}. The massless neutrinos are
described by the conventional Dirac Lagrangian. The resulting model is given by
the coupled Dirac and scalar fields. The grand thermodynamic potential of the
coupled model can be derived from the euclidian functional integral
representation of the grand partition function. The dynamics of the coupled
model is governed by the Friedmann equations.

Throughout the paper we use natural units where $\hbar=c=k_B=1$.
%
%
\subsection{Bosonic Scalar Field}\label{BField}
%
%
%
The bosonic scalar field Hamiltonian in the FLRW metric reads as
\cite{Weinberg08,BirDav}
\begin{equation}
\label{HamBa}
    H_B= \int a^3 d^3 x ~\Big[
    \frac12 \dot{\varphi}^2 +\frac{1}{2 a^2}
    (\nabla \varphi)^2 + U(\varphi) \Big]~,
\end{equation}
where the comoving volume $ V=\int d^3 x$, while the physical
volume $V_{\mathrm{phys}}=a^3(t) V$.
Since this field does not carry a conserved charge (number), the chemical
potential $\mu=0$. The grand partition function in the functional integral
representation:
\begin{equation}
\label{ZB}
    \mathcal{Z}_B \equiv  \mathrm{Tr}  \ \! \mathrm{e}^{-\beta \hat{H}}=
    \int \mathcal{D}\varphi  \ \!
    \mathrm{e}^{-S_B^E}
\end{equation}
with the bosonic euclidian action
\begin{equation}
\label{SEBa}
    S_B^E= \int_0^\beta d \tau \int a(t)^3 d^3 x ~\Big[
    \frac12 (\partial_\tau \varphi)^2 +\frac{1}{2 a^2}
    (\nabla \varphi)^2 + U(\varphi) \Big]~,
\end{equation}
where  $\varphi= \varphi(\mathbf{x}, \tau)$.

It is instructive to find the partition function of the free scalar field
$U(\varphi)= \frac12 M^2_b \varphi^2$ following the methods explained by Kapusta
and Gale \cite{Kapusta} for the case of the Minkowski metric. Rescaling of the
field
\begin{equation}
\label{resc}
    \tilde{\varphi}= a^{3/2}  \varphi
\end{equation}
changes the partition function (\ref{ZB}) by a thermodynamically irrelevant
prefactor.  The functional integration over $\tilde{\varphi}$  of the Gaussian
action  gives
\begin{equation}
\label{LogZBa}
    \log \mathcal{Z}_B= - V \int \frac{d^3 k}{(2 \pi)^3}
    \Big[\beta \sqrt{M^2_b+k^2/a^2}+\log \Big(1-\mathrm{e}^{-\beta \sqrt{M^2_b+k^2/a^2}
   } \Big) \Big]~.
\end{equation}
Then the density (with respect to the physical volume) of the thermodynamic
potential is given by
\begin{eqnarray}
      \Omega_B &\equiv& - \frac{1}{\beta a^3 V} \log \mathcal{Z}_B=-P_B \nonumber \\
      \label{OmegBa}
       &=& \int \frac{d^3 k}{(2 \pi)^3}
    \big[\varepsilon +   \frac{1}{\beta } \log \big(1-\mathrm{e}^{-\beta
    \varepsilon} \big) \big]~,
\end{eqnarray}
where $\varepsilon= \sqrt{M^2_b+k^2}$ and $P_B$ is the pressure due to the bosonic field.
%
%
\subsection{Free Dirac Spinor Field}\label{DField}
%
%
The Dirac Hamiltonian in the FLRW metric is \cite{BirDav}
\begin{equation}
\label{DHa}
    H_D=\int a^3 d^3 x ~\bar{\psi}
     \big(-\frac{\imath}{a}
    \gammab \cdot \nablab +m  \big) \psi~.
\end{equation}
The grand partition function is given by the following Grassmann functional
integral:
\begin{equation}
\label{ZDirac}
    \mathcal{Z}_D \equiv  \mathrm{Tr} \ \! \mathrm{e}^{-\beta(\hat{H}-\mu
    \hat{Q})}=\int \mathcal{D}\bar{\psi}\mathcal{D}\psi \ \!
    \mathrm{e}^{-S_D^E}
\end{equation}
where the conserved charge (lepton number)
operator $\hat{Q}= \int a^3 d^3 x \psi^\dag \psi$ and the euclidian action
\begin{equation}
\label{SEDa}
    S_D^E=\int_0^\beta d \tau \int a(t)^3 d^3 x ~\bar{\psi}(\mathbf{x}, \tau)
     \Big(\gamma^o \frac{\partial}{\partial \tau}-\frac{\imath}{a}
    \gammab \cdot \nablab +m -\mu \gamma^o \Big) \psi(\mathbf{x}, \tau).
\end{equation}
By rescaling the Grassmann fields (\ref{resc}) and using the
standard techniques \cite{Kapusta}, we get the thermodynamic
potential density (pressure) as a function of the chemical
potential and temperature:
\begin{eqnarray}
      \Omega_D &\equiv& - \frac{1}{\beta a^3 V} \log \mathcal{Z}_D=-P_D \nonumber \\
      \label{OmegDa}
       &=&  -2  \int \frac{d^3 k}{(2 \pi)^3}
    \big[\varepsilon +  \frac{1}{\beta } \log \big(1+\mathrm{e}^{-\beta
    \varepsilon_-} \big)+ \frac{1}{\beta } \log \big(1+\mathrm{e}^{-\beta
    \varepsilon_+} \big) \big]~,
\end{eqnarray}
where
\begin{equation}
\label{eps}
  \varepsilon(k) = \sqrt{m^2+k^2}~,
\end{equation}
and $\varepsilon_\pm= \varepsilon(k) \pm \mu$.
The first term on the r.h.s. of Eq.
(\ref{OmegDa}) corresponds to the vacuum contribution to the
thermodynamic potential (pressure):
\begin{equation}
\label{VacD}
    -\Omega_0= P_0 =  2  \int \frac{d^3 k}{(2 \pi)^3}\varepsilon(k)
\end{equation}
Introducing the notation for the Fermi distribution
function
\begin{equation}
\label{nF}
    n_F(x) \equiv \frac{1}{\mathrm{e}^{\beta x}+1}~,
\end{equation}
Eq.~(\ref{OmegDa}) can be brought to the following form:
\begin{equation}
\label{OmegaPDmod}
    -\Omega_D=P_D= P_0+ \frac{1}{3 \pi^2}\int_0^\infty
    \frac{k^4 dk}{\varepsilon(k)}\big[
    n_F(\varepsilon_-)+n_F(\varepsilon_+) \big]
\end{equation}
%
%
\subsection{Coupled Model: Scalar Field and Dirac Massless Fermions}\label{CoupledM}
%
%
Let us consider a scalar bosonic field interacting via a Yukawa
coupling with massless Dirac fermions. The euclidian action of
the model in the FLRW metric reads:
\begin{equation}
\label{Stot}
    \mathcal{S}=S_B^E+S_D^E \big\vert_{m=0}+g \int_0^\beta d \tau \int a^3 d^3 x ~\varphi \bar{\psi}
     \psi
\end{equation}
The path integral for the partition function of the coupled model
is:
\begin{equation}
\label{Ztot}
    \mathcal{Z}=\int \mathcal{D} \varphi \mathcal{D}\bar{\psi}\mathcal{D}\psi \ \!
    \mathrm{e}^{-\mathcal{S}}
\end{equation}
The Grassmann fields can be formally integrated out resulting in
\begin{equation}
\label{Zphi}
    \mathcal{Z}=\int \mathcal{D} \varphi \ \!
    \mathrm{e}^{-\mathcal{S}(\varphi)}=
    \int \mathcal{D} \varphi \ \! \exp \big[-S_B^E+\log \mathrm{Det}
    \hat{D}(\varphi) \big]~,
\end{equation}
where the Dirac operator
\begin{equation}
\label{D}
    \hat{D}(\varphi)=\gamma^o \frac{\partial}{\partial \tau}- \frac{\imath}{a}
    \gammab \cdot \nablab +g \varphi(\mathbf{x}, \tau) -\mu \gamma^o
\end{equation}
The thermodynamic potential $\Omega$ of the model (\ref{Stot}) at
tree level can be found by evaluating the path integral
(\ref{Zphi}) in the saddle-point approximation. Assuming the
existence of a constant $(\mathbf{x}, \tau)$-independent field
$\phi_c$ which minimizes the action $\mathcal{S}(\varphi)$, the
term $\log \mathrm{Det} \hat{D}$ can be evaluated exactly, and
fermionic contribution to the thermodynamic potential is given by
Eqs.~(\ref{OmegDa}) or (\ref{OmegaPDmod}) with the fermionic mass
\begin{equation}
\label{Mass}
    m=g \phi_c~.
\end{equation}
The bosonic contribution to the partition function in this approximation is simply
$
   \mathcal{Z} \propto \exp [ - \beta a^3 V U(\phi_c) ]~.
$
The thermodynamic potential density is given then by
\begin{equation}
\label{OmTree}
    \Omega(\phi_c)= U(\phi_c)+\Omega_D(\phi_c)~.
\end{equation}
Self-consistency of the employed saddle-point approximation
naturally coincides with the condition of minimum of the
thermodynamic potential at equilibrium (at fixed temperature and
chemical potential):
\begin{equation}
\label{SSB}
    \frac{\partial \Omega(\varphi)}{\partial \varphi}
    \Big\vert_{\varphi=\phi_c}=0~,
\end{equation}
and
\begin{equation}
\label{Min}
    \frac{\partial^2 \Omega(\varphi)}{\partial \varphi^2}
    \Big\vert_{\varphi=\phi_c}>0~,
\end{equation}
Note that a non-trivial solution $\phi_c$ of Eq.~(\ref{SSB}) (if it
exists) is called the classical field: it is the average of the
bosonic field, i.e., $\phi_c=\langle \varphi \rangle$.
Eqs.~(\ref{Mass},\ref{OmTree},\ref{SSB}) can be brought to the
equivalent form:
\begin{equation}
\label{AltMin}
    U^\prime (\phi_c)+ g \rho_s=0~,
\end{equation}
where the scalar fermionic density (a.k.a. the chiral density)
$\rho_s$ is given by the following expression:
\begin{equation}
\label{rhos}
    \rho_s \equiv \frac{\langle \hat N \rangle }{V}=
    \frac{ \partial \Omega_D}{\partial m}
    = \rho_0 + \frac{m}{\pi^2}\int_0^\infty
    \frac{k^2 dk}{\varepsilon(k)}\big[
    n_F(\varepsilon_-)+n_F(\varepsilon_+) \big]~,
\end{equation}
and $ \hat N =  \int d^3 x \bar \psi  \psi$. Here  $\rho_0$ stands for the
vacuum contribution to the chiral condensate:
\begin{equation}
\label{rho0}
  \rho_0 \equiv \frac{ \partial \Omega_0}{\partial m}=
  - \frac{m}{\pi^2}\int_0^\infty
    \frac{k^2 dk}{\varepsilon(k)}~.
\end{equation}
Note that even if the time, i.e., $a(t)$, does not enter \textit{explicitly} in
the equations for the thermodynamic quantities of the coupled, fermionic or
bosonic models (\ref{OmegBa},\ref{OmegDa},\ref{OmTree},\ref{AltMin},\ref{rhos}),
and they look like their counterparts in a flat static Universe, such parameters
as, e.g., the temperature and chemical potential in those equations are
time-dependent, i.e., $T=T(a)$ and $\mu= \mu(a)$. The particular form of the
dependencies $T(a)$ and $\mu(a)$ must be determined from the Friedmann continuity
equation (\ref{Frid0}) which relates the energy density $\rho(T)$ and pressure
$P(T)$ to the evolution of $a(t)$\cite{Weinberg08,Dodelson}.
In addition, the fermionic mass $m \propto \phi_c$ in the coupled model is also time
varying, since the time enters into $\phi_c$  (\ref{AltMin}) via $T, \mu$, and all three
functions $m(a)$, $T(a)$ and $\mu(a)$ are governed by the Friedmann
equations (\ref{friedmann1},\ref{friedmann2},\ref{Frid0}).

The present theory works consistently for the physical quantities (bosonic or
fermionic) measured with respect to their vacuum contributions. So, in the rest
of the paper we will employ the thermodynamic quantities with subtracted vacuum
contributions, keeping however, the same notations, e.g.:
\begin{equation}
\label{Vac}
    \Omega_D \mapsto \Omega_D- \Omega_0~,
    P_D \mapsto P_D- P_0~,
    \rho_s \mapsto \rho_s- \rho_0~.
\end{equation}
Then, according to Volovik  \cite{Volovik}, the pressure and energy of the pure
and equilibrium vacuum is exactly zero. (The renormalization of the vacuum
terms is, of course a very subtle issue. There are alternative approaches to
this problem known from the literature. See, e.g., \cite{deVega07,Shapiro2}.)
%
%
%
%
%
\section{Analysis of the Mass (Gap) Equation: General Properties}\label{MasEq}
%
%
%
%
In cases interesting for cosmological applications, the scalar field potential
$U(\varphi)$ does not have a non-trivial minimum, and the generation of the
fermion mass (i.e. a solution of (\ref{SSB}) $0 < \phi_c < \infty$) is due to the
interplay between the scalar and fermionic contributions to the total
thermodynamic potential (\ref{OmTree}).

From now on we adapt our equations for the case of equal number
of fermions and antifermions and $\mu=0$, as discussed in
Sec.~\ref{Outlook}. Keeping in mind the neutrinos, we assume an
extra flavor index of fermions with the number of flavors
$\mathfrak{s}$. (For neutrinos $\mathfrak{s}=3$.) We also assume
the flavor degeneracy of the fermionic sector.

Before proceeding further, we need to make some important observations regarding the
behavior of the coupled model in two limiting cases. Assuming that a
non-trivial solution of (\ref{SSB}) with finite $m$  exists, the fermionic
contribution to the thermodynamic potential (pressure) (\ref{OmegaPDmod}) can
be written as:
\begin{equation}
\label{OmDT}
    -\Omega_D=P_D= \frac{2 \mathfrak{s}}{3 \pi^2 \beta^4}
    \mathcal{I}_p(\beta m) ~,~~\mu=0~,
\end{equation}
where the integral defined as
\begin{equation}
\label{Ip}
    \mathcal{I}_p(\kappa) \equiv
    \int_\kappa^\infty\frac{(z^2 -\kappa^2)^{3/2}}{e^z+1}dz
\end{equation}
can be evaluated analytically in two cases:
\begin{equation}
\label{IpAs}
    \mathcal{I}_p(\kappa) =  \left\{
                \begin{array}{ll}
    \frac{7 \pi^4}{120}- \frac{\pi^2}{8} \kappa^2+\mathcal{O}(\kappa^4)~, & \kappa < 1 \\[0.3cm]
    3 \kappa^2 K_2(\kappa)+\mathcal{O}(e^{-2 \kappa}) ~, & \kappa \gtrsim 1
                \end{array}
       \right.
\end{equation}
where $K_\nu (x)$ is the modified Bessel function of the second
kind.

\textit{In the (classical) low-temperature regime}
\begin{equation}
\label{LT}
    \beta m \equiv   \frac{m}{T}\gg 1
\end{equation}
the above equation results in
\begin{equation}
\label{OmDLT}
    -\Omega_D=P_D= \frac{2 \mathfrak{s} m^2}{\pi^2 \beta^2}
    K_2(\beta m) + \mathcal{O}(e^{-2 \beta m}) ~.
\end{equation}
To leading order
\begin{equation}
\label{OmLTApprox}
    -\Omega_D=P_D \approx \frac{\sqrt{2}\mathfrak{s} }{\pi^{3/2}}
    T (T m)^{3/2} e^{-m/T}~.
\end{equation}
The chiral condensate density (\ref{rhos})
\begin{equation}
\label{Trhos}
    \rho_s =\frac{2 \mathfrak{s} m}{ \pi^2 \beta^2} \int_{\beta m}^\infty
    \frac{(z^2-(\beta m)^2)^{\frac12} }{e^z+1}dz~, ~~\mu=0
\end{equation}
can be also evaluated in the low-temperature limit as
\begin{equation}
\label{TrhosLT}
    \rho_s = \frac{2 \mathfrak{s}  m^2}{\pi^2 \beta}
    K_1(\beta m) + \mathcal{O}(e^{-2 \beta m}) ~,
\end{equation}
which gives to leading order
\begin{equation}
\label{TrhoLTApprox}
     \rho_s  \approx \frac{\sqrt{2}\mathfrak{s}  }{\pi^{3/2}}
    (T m)^{3/2} e^{-m/T}~.
\end{equation}
In this limit the fermions enter the regime of \textit{a classical ideal gas.}
Indeed, the fermionic particle (antiparticle) density
\begin{equation}
\label{npm}
    n_+=n_-  = \frac{\mathfrak{s}}{ \pi^2 \beta^3} \int_{\beta m}^\infty
    \frac{z (z^2-(\beta m)^2)^{\frac12} }{e^z+1}dz
\end{equation}
in the low-temperature limit yields
\begin{equation}
\label{npmLT}
    n_\pm = \frac{ \mathfrak{s} m^2}{\pi^2 \beta}
    K_2(\beta m) + \mathcal{O}(e^{-2 \beta m}) ~,
\end{equation}
and to leading order:
\begin{equation}
\label{npmLTApprox}
    n_\pm  \approx \frac{\mathfrak{s}}{\sqrt{2} \pi^{3/2}}
    (T m)^{3/2} e^{-m/T}~.
\end{equation}
We see from Eqs.~(\ref{OmDLT},\ref{npmLT}) that up to terms
$\mathcal{O}(e^{-2 \beta m})$, the fermions satisfy the ideal gas
equation of state
\begin{equation}
\label{IdGas}
    P_D \approx (n_+ +n_-)T~,
\end{equation}
and the chiral density is equal to the total particle density $n$.
\begin{equation}
\label{densities}
    \rho_s \approx n \equiv n_+ +n_-  ~.
\end{equation}
\textit{In the (ultra-relativistic) high-temperature regime}
\begin{equation}
\label{UR}
    \frac{m}{T} \ll 1
\end{equation}
one obtains
\begin{equation}
\label{OmDUR}
    -\Omega_D=P_D \approx \frac{7 \pi^2 \mathfrak{s} }{180} T^4
    -\frac{\mathfrak{s}}{12}(mT)^2 ~.
\end{equation}
To leading order the chiral condensate is
\begin{equation}
\label{TrhoUR}
    \rho_s  \approx    \frac{\mathfrak{s}}{6} m T^2~,
\end{equation}
while the particle density is
\begin{equation}
\label{npmUR}
    n_\pm  \approx \frac{3 \mathfrak{s} \zeta(3)}{2 \pi^2} T^3~.
\end{equation}

Now we can make some general observations of the fermionic mass generation in
the coupled model:

\textit{(i)} It is obvious from the sign of $\rho_s $ (cf.
\ref{rhos},\ref{Trhos}) that \textit{non-trivial solutions of (\ref{AltMin})
are impossible for a monotonically increasing potential $U(\varphi)$.} That
rules out some popular potentials, e.g.,  $U \propto \log(1+ \varphi/M)$
\cite{Fardon:2003eh,Copeland:2006wr} for this Yukawa-coupling driven scenario
of the mass generation.

\textit{(ii)} The monotonously decreasing slow-rolling DE
potentials (\cite{Ratra:1988,Wetterich:1988} and for reviews, see
\cite{Peebles:2002gy,Copeland:2006wr}), e.g., $U \propto
\varphi^{-\alpha}$ or $U \propto \exp[-A\varphi^{\gamma}]$, do
have \textit{a window of parameters} wherein non-trivial
solutions of (\ref{AltMin}) exist. As we can see from
(\ref{TrhoLTApprox}), for those decreasing potentials the mass
equation (\ref{AltMin}) always has a trivial solution $m=g
\phi_c= \infty$ for the minimum of the thermodynamic potential
(\ref{OmTree}). \footnote{ \label{noteOmFE} Recall  that the
grand thermodynamical potential is equal to the free energy for the
case $\mu=0$. } This solution corresponds to a ``doomsday" vacuum
state \cite{Volovik}, when the Universe reached its true ground
state with zero dark energy density and completely frozen out
fermions. A non-trivial solution of (\ref{AltMin}), corresponding
to another minimum of the potential (\ref{OmTree}), is totally
due to the fermionic contribution. Since the latter freezes out
in the limit $T \to 0$, it is clear qualitatively that such a
solution $0<m< \infty$ can exist only above a certain temperature.
For a more quantitative account of these phenomena we need to
assume some specific form of the DE potential. This will be done
in the following section.

\textit{(iii)} To explain the differences between the present
study and earlier related work on mass varying fermions (see
\cite{Anderson:1997un,Hoffman:2003ru,Fardon:2003eh,Afshordi:2005ym}
and more references there), some clarifications are warranted. It
is usually assumed in the literature that the low-temperature
regime formulas are applicable, and according to
(\ref{densities}) $\rho_s =n$. The approximation for
(\ref{AltMin}) then can be written as $\partial U/
\partial m +n=0$. The latter is interpreted as a result of minimization of some
effective potential $U_{\mathrm{eff}}=U+nm$ with fixed $n$, which always has a
non-trivial minimum $0<m< \infty$ for the class of decreasing potentials $U$,
see, e.g., \cite{Anderson:1997un,Hoffman:2003ru}. It turns out that such an
approximation changes the picture \textit{qualitatively}.

In what follows, we explore in detail the predictions of the consistent  mass
equation (\ref{AltMin}) on the mass varying scenario for the coupled model with
a specific DE potential ansatz.
%
%
%
%
%
\section{Coupled Model with the Ratra-Peebles Quintessence Potential}\label{PRPot}
%
%
\subsection{Mass Equation and Critical Temperature}\label{MTc}
%
%
Now we analyze in detail our coupled model for a particular choice of $U(\varphi)$,
the so-called Ratra-Peebles quintessence potential \cite{Ratra:1988} :
\begin{equation}
\label{RP}
    U(\varphi)=\frac{M^{\alpha+4}}{\varphi^{\alpha}}~,
\end{equation}
where $ \alpha>0$.
It is convenient to introduce the dimensionless parameters
\begin{equation}
\label{DeltaKappa}
    \Delta \equiv  \frac{M}{T}~,~~ \kappa \equiv \frac{g \varphi}{T}~,~
    \Omega_R \equiv \frac{\Omega}{M^4}~.
\end{equation}
Then the mass equation (\ref{AltMin}) can be written as:
\begin{equation}
\label{MassEq}
    \frac{\alpha\pi^2 }{2 \mathfrak{s}} g^\alpha \Delta^{\alpha+4}=\mathcal{I}_\alpha(\kappa)~,
\end{equation}
where we introduced
\begin{equation}
\label{Ia}
    \mathcal{I}_\alpha(\kappa) \equiv
    \kappa^{\alpha+2}\int_\kappa^\infty\frac{\sqrt{z^2 -\kappa^2}}{e^z+1}dz~.
\end{equation}
According to the relation Eq.~(\ref{Mass}) between the fermionic
mass $m$ and the classical field, we get $m=T \kappa_c$, where
$\kappa_c$ is the solution of Eq.~(\ref{MassEq}) corresponding to
the minimum of the thermodynamic potential which reads now as (cf.
Eq.~(\ref{Ip})):
\begin{equation}
\label{OmR}
    \Omega_R= g^\alpha \Big(\frac{\Delta}{\kappa} \Big)^\alpha
    -\frac{2}{3 \pi^2} \frac{1}{\Delta^4} \mathcal{I}_p(\kappa)~.
\end{equation}
The dimensionless Yukawa coupling constant $g \sim 1$. To reduce the number
of model parameters we can set $g=1$. This is equivalent to the simultaneous
rescaling $g \varphi \mapsto \tilde \varphi $ and
$M g^\frac{\alpha}{\alpha+4} \mapsto \tilde M$. \footnote{ \label{noteG}
One can check this scaling also holds for the dynamics  of the model,
considered in Section~\ref{Dynamics}. In particular,
the neutrino masses do not depend on the value of $g$. To avoid cluttering
of notations we will drop tildes in the rescaled parameters.}
For simplicity, we also restrict the number
of flavors $\mathfrak{s}=1$.

We define the mass of the scalar field as:
\begin{equation}
\label{mphi}
    m_\phi^2=  \frac{\partial^2 U(\varphi)}{\partial \varphi^2}
    \Big\vert_{\varphi=\phi_c}~.
\end{equation}
In terms of the dimensionless parameters it reads
\begin{equation}
\label{mphi2}
    \frac{m_\phi}{M}=\sqrt{\alpha(\alpha+1)} \Big(\frac{\Delta}{\kappa_c} \Big)^{\frac{\alpha+2}{2}}
\end{equation}

\begin{figure}[h]
\epsfig{file=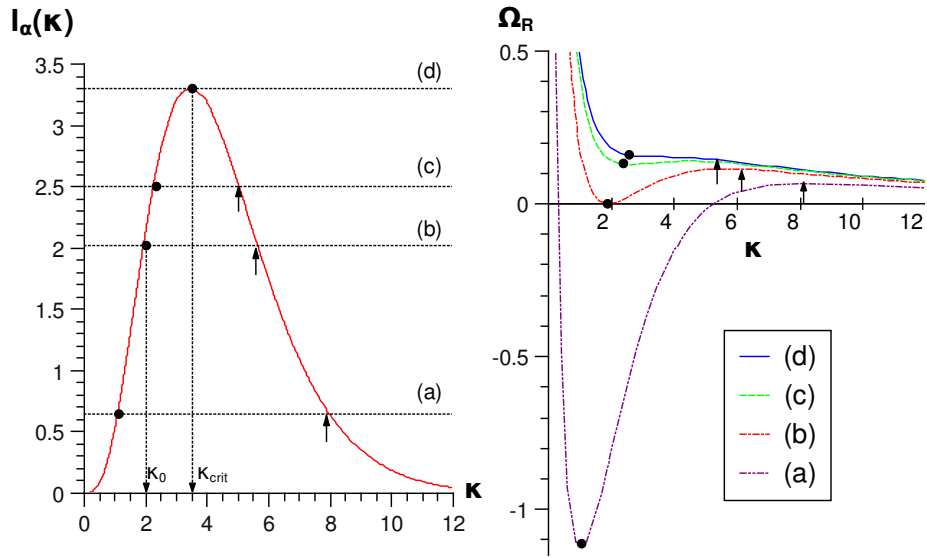,width=0.55\textwidth,angle=270}
\caption{(Color online) Left: Graphical solutions of the mass
equation (\ref{MassEq}) for different  values of $\Delta \equiv
M/T$ ($\alpha=1$). Right: dimensionless density of the
thermodynamic potential (\ref{OmR}). The thermodynamically stable
solutions of Eq.~(\ref{MassEq}) indicated by the large dots
correspond to the minima of the potential. The arrows indicate
the unstable solutions of the mass equation, corresponding to the
maxima of the potential. } \label{IaOm}
\end{figure}

It is important to realize that the integral $\mathcal{I}_\alpha(\kappa)$ on the
r.h.s. of the mass equation is bounded. The quantitative parameters of the function
$\mathcal{I}_\alpha(\kappa)$ depend on $\alpha$, but its shape is always similar to the curve
shown in Fig.~\ref{IaOm} for $ \alpha=1$. So, there exists a maximal $\Delta_{\mathrm{crit}}$
(critical temperature $T_{\mathrm{crit}}$) such that for $\Delta> \Delta_{\mathrm{crit}}$
($T< T_{\mathrm{crit}}$) only a trivial solution $m= \infty$ exists, and the stable vacuum
has zero energy and pressure.
%
%

The mass equation Eq.~(\ref{MassEq}) is solved numerically for
various values of its parameters, and the characteristic results
are shown in Fig.~\ref{IaOm}. The numerical results can be
complemented by an approximate analytical treatment of the
problem. The latter turns out to be quite accurate and greatly helps
 in gaining intuitive understanding of the results.

It is easy to evaluate $\mathcal{I}_\alpha(\kappa)$ to leading order:
\begin{equation}
\label{IaAs}
    \mathcal{I}_\alpha(\kappa) \approx  \left\{
                \begin{array}{ll}
    \frac{\pi^2}{12} \kappa^{\alpha+2}~, & \kappa < 1 \\[0.3cm]
    \kappa^{\alpha+3} K_1(\kappa)~, & \kappa \gtrsim 1
                \end{array}
       \right.
\end{equation}
For the critical point where
$\mathcal{I}_\alpha^{\phantom{.}\prime} (\kappa_{\mathrm{crit}}) =0$, we obtain:
\begin{eqnarray}
\label{kapmax}
&~&  \kappa_{\mathrm{crit}}  \approx \nu~, ~~\nu \equiv \alpha+ \frac52~; \\
\label{Iamax}
 &~& \mathcal{I}_\alpha (\kappa_{\mathrm{crit}})  \approx
  \sqrt{\frac{\pi}{2}} \nu^\nu e^{-\nu}.
\end{eqnarray}

The most important conclusion we draw from Fig.~\ref{IaOm} is that
\textit{there are three phases in the model's phase diagram.} We analyze each
of them in the following subsections.
%
%
\subsubsection{Stable (massive) phase: $\Delta < \Delta_\circ~(T_\circ <T < \infty)$}\label{Stable}
%
%
In this range of parameters the equation (\ref{MassEq}) has two nontrivial solutions.
The root $\kappa_c< \kappa_\circ$ indicated with a large dot in Fig.~\ref{IaOm}
(case a) gives  the fermionic mass and corresponds to a global minimum of the potential. So
it is a thermodynamically \textit{stable state}. In this phase $\Omega(\kappa_c) <0$,
so the pressure is positive $P>0$. Another non-trivial root of (\ref{MassEq}) corresponds
to a thermodynamically \textit{unstable state}
(maximum of $\Omega$ indicated with an arrow in Fig.~\ref{IaOm}).
There is a trivial third root of the mass equation $\kappa = \infty$. At these
temperatures it corresponds to the \textit{metastable vacuum state} $\Omega=0$.

\begin{figure}[h]
\epsfig{file=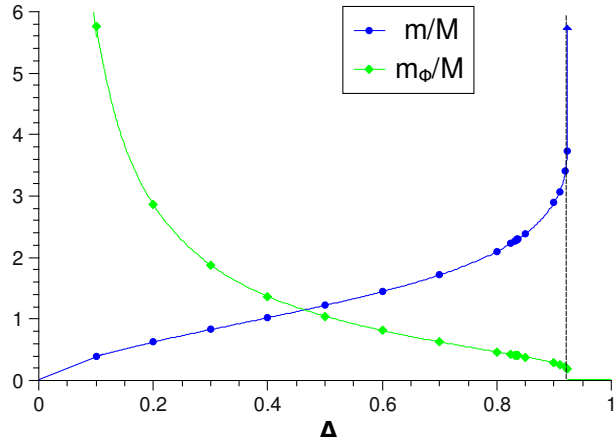,width=0.40\textwidth,angle=270} \caption{(Color
online) Masses of the fermionic and scalar fields ($m$ and $m_\phi$ resp.)
as functions of $\Delta \equiv M/T$, $\alpha =1$. At
$\Delta > \Delta_{\mathrm{crit}}~(T <T_{\mathrm{crit}})$ the stable phase corresponds
to $m = \infty$ and $m_\phi =0$ } \label{Masses}
\end{figure}

In the high-temperature region of this phase where $\Delta \ll 1$
the fermionic mass is small (see Fig.~\ref{Masses}):
\begin{equation}
\label{msmall}
    \frac{m}{M} \approx \Big( \sqrt{6 \alpha} \frac{M}{T}  \Big)^{\frac{2}{\alpha+2}}
    \propto T^{-\frac{2}{\alpha+2}}
\end{equation}
The fermionic contribution to the thermodynamic potential is dominant, and
it behaves to leading order as the potential of the ultra-relativistic fermion gas
(cf. Eq.~(\ref{OmDUR})):
\begin{equation}
\label{OmUR}
    \Omega=-P = -\frac{7 \pi^2 }{180} T^4+ \mathcal{O}(T^{ \frac{2 \alpha}{\alpha+2}}) ~.
\end{equation}
One can check that the subleading term in the above expression combines the
DE potential contribution and the first fermionic mass correction,
which are both of the same order.
%
%

It is important to stress that in this coupled model with the
slow-rolling potential Eq.~(\ref{RP}), the mass generation does
not follow a conventional Landau thermal phase transition
scenario. There is no critical temperature below which the
chiral symmetry is spontaneously broken and the mass is generated.
Instead the mass grows smoothly as $\kappa_c \propto
\Delta^{\frac{\alpha+4}{\alpha+2}}$, albeit starting from the
``point" $T= \infty$. From physical grounds we expect the
applicability of the model to have the upper temperature bound:
\begin{equation}
\label{upper T}
    T \lesssim T_{\mathrm{RD}}~,
\end{equation}
where $T_{\mathrm{RD}}$ is roughly the temperature of the boundary between inflation
and the radiation-dominated era. The high-temperature result (\ref{OmUR}) shows that the
stable massive phase of the present model can indeed be extended up to those temperatures.

The scalar field and  fermionic masses demonstrate opposite temperature dependencies.
The scalar field is ``heavy" at high temperatures:
\begin{equation}
\label{mphiUP}
    m_\phi \approx \sqrt{\frac{\alpha+1}{6}} T~,~~ \Delta \ll 1~,
\end{equation}
however its mass decreases together with the temperature. In contrary, the fermionic
mass $m$ monotonously increases with decreasing temperature. The exact numerical results
for the two masses are shown in Fig.~\ref{Masses}
%
%
\subsubsection{Metastable (massive) phase: $\Delta_\circ <\Delta < \Delta_{\mathrm{crit}}
~(T_{\mathrm{crit}} <T < T_\circ $)}\label{Metastable}
%
%
Upon increasing $\Delta$ we reach a certain value $\Delta_\circ$ corresponding to
a critical temperature $T_\circ$ when the thermodynamic potential has two
degenerate minima $\Omega(\kappa_\circ)=P(\kappa_\circ)=\Omega(\infty)=0$.
This is shown in Fig.~\ref{IaOm} (case b). After this point, when the temperature decreases
further in the range $\Delta_\circ <\Delta < \Delta_{\mathrm{crit}}$ (here
$\Delta_{\mathrm{crit}}$ stands for the maximal value of $\Delta$ when a non-trivial
solution of the gap equation (\ref{MassEq}) exists, see  Fig.~\ref{IaOm}),
the two minima of the thermodynamic potential exchange their roles. The root $\kappa_c$
now  becomes a metastable state with $\Omega(\kappa_c)>0$, i.e., with the negative pressure
$P(\kappa_c)<0$, while the stable state of the system corresponds to the true stable vacuum
of the Universe \cite{Volovik} $\Omega(\infty)=P(\infty)=0$. See Fig.~\ref{IaOm} (case c).
The system's state in the local minimum $\Omega(\kappa_c)$ is analogous to a metastable
supercooled liquid. We disregard the exponentially small probability of tunneling of the
fermions from the metastable state $\Omega(\kappa_c)$ into the vacuum state
$\Omega(\infty)=0$\cite{Linde}. Accordingly, the fermionic mass in this phase is determined
by the root $\kappa_c$ of (\ref{MassEq}).

In the metastable phase $\kappa_c \gtrsim 1$, so by using Eqs.~(\ref{OmR},\ref{IpAs},\ref{MassEq})
we obtain the potential:
\begin{equation}
\label{OmRmet}
    \Omega_R \approx \Big(\frac{\Delta}{\kappa_c} \Big)^\alpha
    \Big\{ 1- \frac{\alpha}{\kappa_c} -\frac{3 \alpha}{2 \kappa_c^2} \Big\}~.
\end{equation}
From the above result we can find the metastability point $\Omega(\kappa_\circ)=0$ as
\begin{equation}
\label{kappa0}
    \kappa_\circ \approx \frac{\alpha}{2} \Big( 1+\sqrt{1+\frac{6}{\alpha}} \Big)
\end{equation}
Expanding $\mathcal{I}_\alpha(\kappa)$ near its maximum and using
Eqs.~(\ref{IaAs},\ref{kapmax},\ref{Iamax}) along with the gap
equation Eq.~(\ref{MassEq}), we obtain the following  equation:
\begin{equation}
\label{mDelta}
    \frac{(\kappa_c- \kappa_{\mathrm{crit}})^2}{2 \nu}
    \approx 1-\Big(\frac{\Delta}{\Delta_{\mathrm{crit}}} \Big)^{\alpha+4}~.
\end{equation}
On finds from the above equation, e.g., how the mass approaches its critical value:
\begin{equation}
\label{dm}
    m_{\mathrm{crit}}-m \propto
    \Big(\frac{T}{T_{\mathrm{crit}}} -1  \Big)^{1/2 }~,
\end{equation}
or the ratios of temperatures and masses at the metastable and critical points.
These latter parameters are given in Table~\ref{CritParameters}.

\begin{table}[h]
  \caption{Masses, critical temperatures and potentials for various values of  $\alpha$.
  All the parameters used in this table are defined in the text.}
  ~\\
\begin{tabular}{|c|c|c|c|c|c|c|c|c|c|}
  \hline
  ~ & ~ & ~ & ~ & ~ & ~ & ~ & ~ & ~ & ~\\[-0.4cm]
  $~~\alpha~~$ & $~~\frac{T_{\mathrm{crit}}}{T_\circ}~~$ & $~~\Delta_{\mathrm{crit}}~~$
  & $~~\frac{m_\circ}{m _{\mathrm{crit}}}~~$ & $~~\frac{m _{\mathrm{crit}}}{M}~~$ & $~~\frac{m^{\mathrm{crit}}_\phi}{M}~~$
  & $~~\frac{T_{\mathrm{crit}}}{M}~~$ & $~~\frac{\Omega_{\mathrm{crit}}}{M^4}~~$ & $~~\frac{\rho_{\mathrm{crit}}}{M^4}~~$
  & $~~w(T_{\mathrm{crit}})~~$ \\
  ~ & ~ & ~ & ~ & ~ & ~ & ~ & ~ & ~ & ~\\[-0.4cm]
  \hline
  1 & 0.90 & 0.91 & 0.558 & 3.86 & 0.187               & 1.10   &0.15              &0.84             &-0.18          \\
  \hline
  2 & 0.95 & 1.04 & 0.70 & 4.35 & 0.130                & 0.97   &0.02              &0.25             &-0.09            \\
  \hline
  4 & 0.98 & 1.44 & 0.81 & 4.52 & 0.048                & 0.70 &$6 \cdot 10^{-4}$   &0.02             &-0.03             \\
  \hline
  10 & 0.99 & 3.00 & 0.91 & 4.16 & $2 \cdot 10^{-3}$   &0.33  &$7 \cdot 10^{-8}$  &$9 \cdot 10^{-6}$ &-0.008            \\
  \hline
\end{tabular}
 \label{CritParameters}
\end{table}
%
%
\subsubsection{Critical point: $\Delta= \Delta_{\mathrm{crit}}
~(T= T_{\mathrm{crit}}$) and phase transition}\label{Crit}
%
%
The critical point of the model corresponds to the case when the
two roots of the mass equation Eq.~(\ref{MassEq}) merge, and the
minimum of the potential disappears. One can check that instead
of the minimum this is an inflection point of the the potential,
i.e., $\Omega_R''(\kappa_{\mathrm{crit}})=0$. This situation is
shown in Fig.~\ref{IaOm} (case d). At this point the system is in the \textit{unstable} state with the fermionic mass
\begin{equation}
\label{mcr}
    \frac{m_{\mathrm{crit}}}{T_{\mathrm{crit}}}=\kappa_{\mathrm{crit}} \approx \nu~.
\end{equation}
In particular, this implies that the fermions are
non-relativistic at the critical temperature. From
Eqs.~(\ref{Iamax},\ref{MassEq}) we find the critical parameter
(see Table~\ref{CritParameters} for its numerical values)
\begin{equation}
\label{Dcr}
    \Delta_{\mathrm{crit}} \approx
    \Big( \frac{\sqrt{2}}{\alpha \pi^{3/2}} \nu^\nu e^{-\nu} \Big)^{\frac{1}{\alpha+4}}~,
\end{equation}
which allows us to evaluate the critical temperature
\begin{equation}
\label{Tcr}
   T_{\mathrm{crit}}=  \frac{M}{ \Delta_{\mathrm{crit}}} ~.
\end{equation}
We can also find the potential at $T_{\mathrm{crit}}$:
\begin{equation}
\label{Omcr}
    \Omega_{\mathrm{crit}} \approx \frac{5 }{2 \nu } \Big(\frac{ \Delta_{\mathrm{crit}} }{\nu} \Big)^\alpha M^4
\end{equation}
Thus, from the viewpoint of \textit{equilibrium thermodynamics}
at $T=T_{\mathrm{crit}}$ the model must undergo a
\textit{first-order (discontinuous) phase transition} and reach
its third \textit{thermodynamically stable (at
$T<T_{\mathrm{crit}}$) phase} corresponding to the vacuum
$\Omega(\kappa= \infty)=P(\kappa= \infty)=0$. During this
transition the fermionic mass given at the critical point by
Eq.~(\ref{mcr}) and the scalar field mass
\begin{equation}
\label{mphicr}
    m^{\mathrm{crit}}_\phi \approx
    \sqrt{\alpha(\alpha+1)} \Big(\frac{\Delta_{\mathrm{crit}}}{\nu} \Big)^{\frac{\alpha+2}{2}}M
\end{equation}
both jump to their values in the vacuum state $m=\infty$ and $m_\phi=0$. See Fig.~\ref{Masses}.

However, the above arguments are based on the minimization of the thermodynamic
potential (i.e. maximization of entropy) at equilibrium. To address the
question of how such a system behaves as the Universe evolves towards the new
equilibrium vacuum state, we need to analyze the dynamics of this phase
transition. More qualitatively, we need to study how the particle at
the point $\kappa_{\mathrm{crit}}$ at the critical temperature (see
Fig.~\ref{IaOm}) rolls down towards its equilibrium at infinity. This issue
will be addressed in Section~\ref{Dynamics}.
%
%
\subsection{Equation of State}\label{rho&w}
%
%
We define the equation of state in the standard form:
\begin{equation}
\label{wdef}
    P=w \rho~,
\end{equation}
where the total pressure in this model is obtained from
Eq.~(\ref{OmR}), while the total energy density ($\rho$) and its
dimensionless counterpart ($\rho_R$) are determined by the
following equation:
\begin{equation}
\label{rho}
    \rho_R \equiv \frac{\rho}{M^4}=\Big(\frac{\Delta}{\kappa} \Big)^\alpha
    +\frac{2}{\pi^2} \frac{1}{\Delta^4} \mathcal{I}_\varepsilon(\kappa)~.
\end{equation}
Here we define the integral
\begin{equation}
\label{Ieps}
    \mathcal{I}_\varepsilon(\kappa) \equiv
    \int_\kappa^\infty \frac{z^2 \sqrt{z^2 -\kappa^2}}{e^z+1}dz~,
\end{equation}
which can be evaluated in two limits of our interest:
\begin{equation}
\label{IepsAs}
    \mathcal{I}_\varepsilon(\kappa) =  \left\{
                \begin{array}{ll}
    \frac{7 \pi^4}{120}- \frac{\pi^2}{24} \kappa^2+\mathcal{O}(\kappa^4)~, & \kappa < 1 \\[0.3cm]
    3 \kappa^2 K_2(\kappa)+\kappa^3 K_1(\kappa)+ \mathcal{O}(e^{-2 \kappa}) ~, & \kappa \gtrsim 1
                \end{array}
       \right.
\end{equation}
In the high-temperature region of the stable massive phase where $\Delta \ll 1$,
the fermionic contribution is dominant, and the energy density
to leading order is that of the ultra-relativistic fermion gas (cf. Eq.~(\ref{OmUR}))
\begin{equation}
\label{rhoUR}
    \rho=\frac{7 \pi^2 }{60} T^4+ \mathcal{O}(T^{ \frac{2 \alpha}{\alpha+2}}) ~.
\end{equation}
Thus, in this regime the model follows approximately the equation of state of a
relativistic gas with $w \approx \frac13$.

\begin{figure}[h]
\epsfig{file=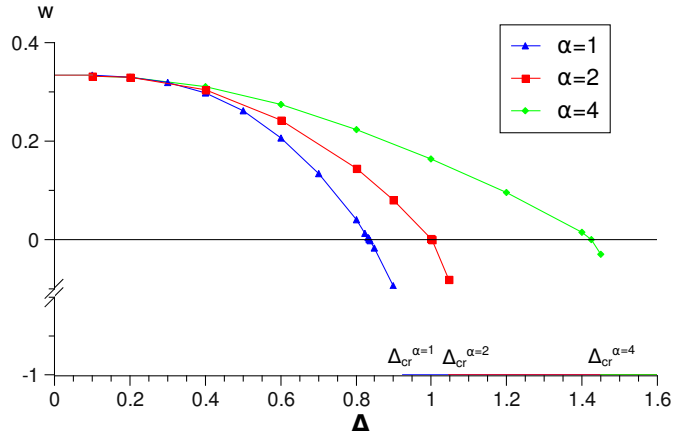,width=0.40\textwidth,angle=270} \caption{(Color online) $w
\equiv P/\rho$ for several values of $\alpha$.   At $\Delta >
\Delta_{\mathrm{crit}}(\alpha)$, i.e., $T <T_{\mathrm{crit}}(\alpha)$ the
equilibrium value $w=-1$ exactly.  } \label{w}
\end{figure}

In the region $\kappa_c \gtrsim 1$ which includes the metastable phase and the
critical point, we obtain by using Eqs.~(\ref{rho},\ref{IepsAs},\ref{MassEq},\ref{OmRmet}):
\begin{equation}
\label{rhoRmet}
    \rho \approx \Big(\frac{\Delta}{\kappa_c} \Big)^\alpha
    \Big\{ 1+\alpha+ \frac{3 \alpha}{\kappa_c} +\frac{9 \alpha}{2 \kappa_c^2} \Big\}~,
\end{equation}
and
\begin{equation}
\label{wmet}
    w \approx -
    \frac{ 1- \frac{ \alpha}{\kappa_c} -\frac{3 \alpha}{2 \kappa_c^2} }{
    1+\alpha+ \frac{3 \alpha}{\kappa_c} +\frac{9 \alpha}{2 \kappa_c^2} }
\end{equation}
The last equation follows very closely the results of the exact numerical calculations
shown in Fig.~\ref{w}.
%
%
At the critical point we evaluate
\begin{equation}
\label{rhocrit}
    \rho_{\mathrm{crit}} \approx \Big(\frac{\Delta_{\mathrm{crit}}}{\nu} \Big)^\alpha
    \Big\{ 1+\alpha+ \frac{3 \alpha}{\nu}  \Big\} M^4~,
\end{equation}
and making a rough estimate, we get a lower bound:
\begin{equation}
\label{wmin}
     w \approx - \frac52 \frac{1}{\nu (1+\alpha+ \frac{3 \alpha}{\nu})}
     \geq -\frac14,~~ \forall ~\alpha \geq 1~.
\end{equation}
Thus for any power law $\alpha \geq 1$, the parameter $w$ of this model at
equilibrium cannot cross the bound $w < -\frac13$, necessary for accelerating
expansion of the Universe $\ddot{a}>0$. \footnote{The relation (\ref{wmin})
$w(T_{\mathrm{crit}})\gtrsim -\frac14$ holds  for the model which contains only
the DE-neutrino coupled fluid. In a more realistic model for the Universe,
baryons and DM also contribute to the total energy density, and as a
consequence $w(T_{\mathrm{crit}})$ increases, see Sec.~\ref{Dynamics}.}

At $T< T_{\mathrm{crit}}$ we obtain the equilibrium value of $w$
in the stable vacuum state from Eqs.~(\ref{OmR},\ref{rho}):
\begin{equation}
\label{wvac}
    w= \lim_{\kappa \to \infty} \frac{P(\kappa)}{\rho(\kappa)}=-1~.
\end{equation}
So the true vacuum in this model corresponds to the Universe
with a cosmological constant in the limit $\Lambda \to 0$.
%
%

\subsection{Speed of Sound}\label{Cs}
%
%
We define the sound velocity as
\begin{equation}
\label{cs2def}
    c_s^2 = \frac{\frac{d P}{d t}}{\frac{d \rho}{d t}}=
    \frac{\frac{d P}{d \Delta}}{\frac{d \rho}{d \Delta}}~,
\end{equation}
where to obtain the second expression we used the fact that the time enters our
formulas only through the temperature $T(a(t))$, so
\begin{equation}
\label{chain}
    \frac{d}{d t}=  \frac{d \Delta}{d t}  \frac{d }{d \Delta}~.
\end{equation}
Let us first consider the temperatures $T \geq T_{\mathrm{crit}}$, i.e.,
$\Delta \leq \Delta_{\mathrm{crit}}$. Then
\begin{equation}
\label{drho}
    \frac{d \rho}{d \Delta}=\frac{\partial \rho}{ \partial \Delta}+
    \frac{\partial \rho}{ \partial \kappa} \cdot \frac{d \kappa}{ d \Delta}
    \Big\vert_{\kappa=\kappa_c}~,
\end{equation}
where $\kappa$ is related to $\Delta$ through the gap equation (\ref{MassEq}):
\begin{equation}
\label{dkap}
    \frac{d \kappa}{ d \Delta} \Big\vert_{\kappa=\kappa_c} \equiv
    \dot{\kappa}_c =\frac{\alpha+4}{\Delta}
    \frac{\mathcal{I}_\alpha(\kappa_c)}{\mathcal{I}_\alpha^{\phantom{.}\prime}(\kappa_c)}=
    \frac{\alpha+4}{\Delta}
    \Big( \frac{d \log \mathcal{I}_\alpha(\kappa_c)}{d \kappa} \Big)^{-1}~.
\end{equation}
Note that for the pressure the following relation
\begin{equation}
\label{dP}
    \frac{d P}{d \Delta}=\frac{\partial P}{ \partial \Delta}~
\end{equation}
holds, since
\begin{equation}
\label{dP0}
    \frac{\partial P}{ \partial \kappa}
    \Big\vert_{\kappa=\kappa_c}=0
\end{equation}
is just another form of the gap equation (\ref{SSB}). Thus
\begin{equation}
\label{cs2Eq}
    c_s^2 = \frac{\frac{\partial P}{ \partial \Delta}}{\frac{\partial \rho}{ \partial \Delta}+
    \frac{\partial \rho}{ \partial \kappa} \dot{\kappa}_c }\Bigg\vert_{\kappa=\kappa_c}~.
\end{equation}
In the high-temperature regime $\Delta \ll 1$ ($\kappa_c \ll 1$),
it is even easier to use the explicit asymptotic expansions for
$P(\Delta)$ and $\rho(\Delta)$ in the definition (\ref{cs2def})
instead of the above formula (\ref{cs2Eq}). A straightforward
calculation gives the result
\begin{equation}
\label{cs2UR}
    c_s^2 \approx \frac13 -b \Delta^{\frac{2(\alpha+4)}{\alpha +2}}~, ~~b>0~,
\end{equation}
consistent with the earlier observation that for $\Delta \ll 1$ the model behaves
as an ultra-relativistic Fermi gas.

In the case $\kappa_c \gtrsim 1$ we find
\begin{equation}
\label{dkapmeta}
    \dot{\kappa}_c \approx \frac{\alpha+4}{\Delta} \frac{\kappa_c}{\nu - \kappa_c}~,
\end{equation}
and
\begin{equation}
\label{cs2meta}
    c_s^2 \approx \frac{\nu - \kappa_c}{\alpha(\alpha+4)(1+\frac{4}{\alpha \nu})}~.
\end{equation}
Everywhere at $T> T_{\mathrm{crit}}$, including the stable and metastable massive
phases  $c_s^2>0$, so the model is stable with respect to the density fluctuations.
The sound velocity vanishes in the limit $T \to T_{\mathrm{crit}}^+$ as
\begin{equation}
\label{cslimit}
    c_s \propto   \sqrt{\nu - \kappa_c}~ \to 0~.
\end{equation}
Qualitatively, the vanishing speed of sound is due to divergent $\dot{\kappa}_c$
(\ref{dkap},\ref{dkapmeta}) at the critical point.

\begin{figure}[h]
\epsfig{file=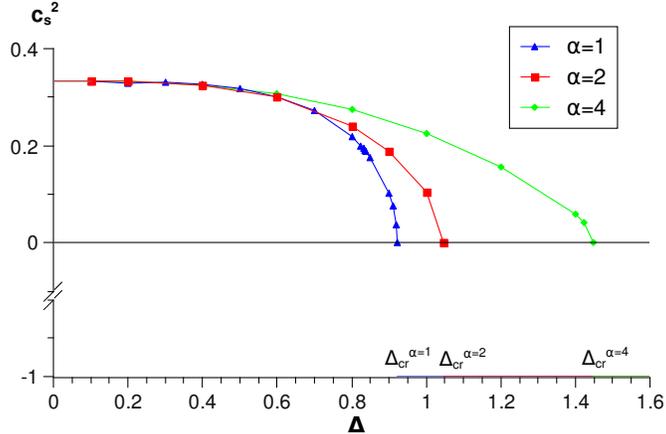,width=0.40\textwidth,angle=270} \caption{(Color
online) The square of the sound velocity for several values of $\alpha$.   At
$\Delta > \Delta_{\mathrm{crit}}(\alpha)$, i.e., $T <T_{\mathrm{crit}}(\alpha)$
the equilibrium value $c_s^2=-1$ exactly.  }\label{sound}
\end{figure}

The above analytical results are in excellent agreement with the numerical
calculations of $c_s^2$ from the formula (\ref{cs2Eq}) shown in Fig.~\ref{sound}.
%
%
%
At the temperatures $T < T_{\mathrm{crit}}$ there is no gap equation relating
$\kappa$ and $\Delta$, so the sound velocity is easily calculated to yield
the value in the equilibrium vacuum state:
\begin{equation}
\label{cs2Vac}
    c_s^2 = \lim_{\kappa \to \infty}\frac{\frac{\partial P}{
    \partial \Delta}}{\frac{\partial \rho}{ \partial \Delta}}=-1~.
\end{equation}
That what is expected for a barotropic perfect liquid with a constant $w$,
where $c_s^2=w$.

%
%
\section{Dynamics of the Coupled Model and Observable Universe}\label{Dynamics}
%
%
%
%
\subsection{Scales and Observable Universe}\label{Scales}
%
%
In order to make a connection between the above model results and the observable
Universe, we need to first conclude where we are now with respect to the
critical temperatures $T_\circ$ and $T_{\mathrm{crit}}$. As one can see from
Table~\ref{CritParameters} for $\alpha \sim 1$, the model has $T_\circ \sim
T_{\mathrm{crit}} \sim M$. We identify the current equilibrium temperature of
the Universe with the cosmic background radiation temperature $T = 2.275~K=2.4
\cdot 10^{-4} ~$eV. Then we see right away that we cannot be above the critical
temperature of the coupled model, since: \\
(i) assumption $T >T_{\mathrm{crit}}$ leads to $M \lesssim 10^{-4} ~$eV, which
in turn implies too small densities  $\rho \sim M^4 \sim 10^{-16} ~\mathrm{eV}^4$, i.e,
four orders of magnitude less than the observable density; \\
(ii) At $T >T_{\mathrm{crit}} $ the equation of state has $w>- \frac14$ (see
Fig.~\ref{w}), which is not even enough to get a positive acceleration
$\ddot{a}>0$, while the observable value $w \approx -1$. \cite{Copeland:2006wr}

So, the first qualitative conclusion is that we are currently below the
critical temperature. The Universe has already passed the stable and metastable
phases and is now unstable, i.e. it is in the transition toward the stable
``doomsday'' vacuum $m = \infty$ and $\Omega =0$.

Since at the temperature of metastability $P_\circ=0$, the transition
occurs somewhere between the beginning of the matter-dominated era ($T_{MD}
\approx 16500~K \approx 1.42 ~\mathrm{eV})$ and now, i.e., $1.4~\mathrm{eV}  \gtrsim
T_{\mathrm{crit}} > T_{\mathrm{now}} \sim 2.4 \cdot 10^{-4} ~\mathrm{eV}$. Because of
Eq.(\ref{Tcr}) this inequality gives us the possible range of the model's single parameter $M$:
\begin{equation}
\label{MT}
    2.4 \cdot 10^{-4} ~\mathrm{eV}~< M \lesssim ~1.4~\mathrm{eV}.
\end{equation}
As we will show in the following, other consistency checks of the model bring
the upper bound of $M$ much lower.
%
%
\subsection{Universe Before the Phase Transition}\label{Before}
\begin{figure}[h]
\epsfig{file=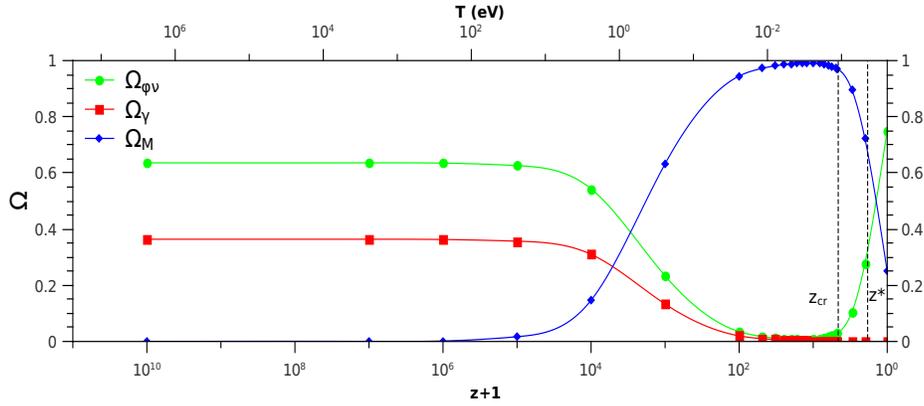,width=0.75\textwidth, height=0.25\textheight}
\caption{(Color online)
Relative energy densities plotted up to the current redshift (temperature, upper axis):
$\Omega_{\varphi\nu}$ -- coupled DE and neutrino contribution; $\Omega_{\gamma}$ --
radiation; $\Omega_{M}$ -- combined baryonic and dark matters. Parameter
$M=2.39 \cdot 10^{-3}$~eV ($\alpha=0.01$), chosen to fit the current densities,
determines the critical point of the phase transition $z_{\mathrm{cr}} \approx 3.67$.
The crossover redshift $z^* \approx 0.83$ corresponds to the  point where the
Universe starts its accelerating expansion.}\label{Omega}
\end{figure}

%
%
In order to apply the results of the coupled model for the calculation of the
parameters of the observable Universe, we need to incorporate the matter (we
will just add up the dark and conventional baryonic matter together) and the
radiation. Assuming a spatially flat Universe, the total energy density is critical, so
\begin{equation}
\label{rhocr1}
 \rho_{\mathrm{tot}}= \rho_{\gamma,\mathrm{now}}/ a^4
+\rho_{M,\mathrm{now}}/a^3 + \rho_{\varphi \nu}(\Delta)= \rho_{\mathrm{cr}} = \frac{3H^2}{8\pi G} ~,
\end{equation}
where from now on we denote $\rho_{\varphi \nu}$ the energy density of the
coupled model given by Eq.~(\ref{rho}). To relate our model's parameters to the
standard cosmological notations, we assume that the temperature is evolving as
that of the blackbody radiation, i.e., $T=T_{\mathrm{now}}/a$. Then
\begin{equation}
\label{Tza}
    \Delta \equiv \frac{M}{T}=\frac{Ma}{T_{\mathrm{now}}}=
    \frac{M}{T_{\mathrm{now}}(1+z)}~.
\end{equation}
We know that
\begin{equation}
\label{rhogamma}
    \rho_\gamma = \frac{\pi^2}{15}T^4~,
\end{equation}
and we set the current density of the coupled scalar field to
the observable value of the dark energy, i.e.,  $3/4$ of the critical density:
\begin{equation}
\label{DEnow}
    \rho_{\varphi\nu, \mathrm{now}}=\frac34 \cdot \frac{3 H_0^2}{8 \pi G}
    \approx 31 \cdot (10^{-3} ~\mathrm{eV})^4~,
\end{equation}
and
\begin{equation}
\label{Mnow}
    \rho_{M, \mathrm{now}} \approx \frac14 \cdot \frac{3 H_0^2}{8 \pi G}~.
\end{equation}
The equations above allow us to plot the relative energy densities
\begin{equation}
\label{Omegas}
    \Omega_\# \equiv \rho_\#  /\rho_{\mathrm{tot}}
\end{equation}
as functions of redshift (or temperature) up to the critical point, see
Fig.~\ref{Omega}. \footnote{We apologize for some abuse of notations, but
using the same Greek letter for the grand thermodynamic potential and relative
densities seems to be standard now. Since these quantities are mainly discussed
in different sections of the paper, we hope the reader will not be confused.}
%
%
%
In the high-temperature limit, the matter term is sub-leading and
\begin{equation}
\label{rhoHT}
  \rho_{\mathrm{tot}}  \approx \rho_\gamma + \rho_{\varphi \nu}
 \approx \frac{\pi^2}{15} \big(1+\frac74 \big) T^4~.
\end{equation}
In this limit, then
\begin{equation}
\label{OmHT}
  \Omega_{\varphi\nu}=\frac{7}{11}\approx 0.636~,~~~
  \Omega_\gamma = \frac{4}{11} \approx 0.363~,
\end{equation}
which agrees well with the numerical results displayed in Fig.~\ref{Omega}.
At the critical point the matter strongly dominates and
$\rho_{\mathrm{M}}/ \rho_{\mathrm{\gamma,  \varphi \nu}} \gtrsim 10^2$.

\begin{figure}[h]
\epsfig{file=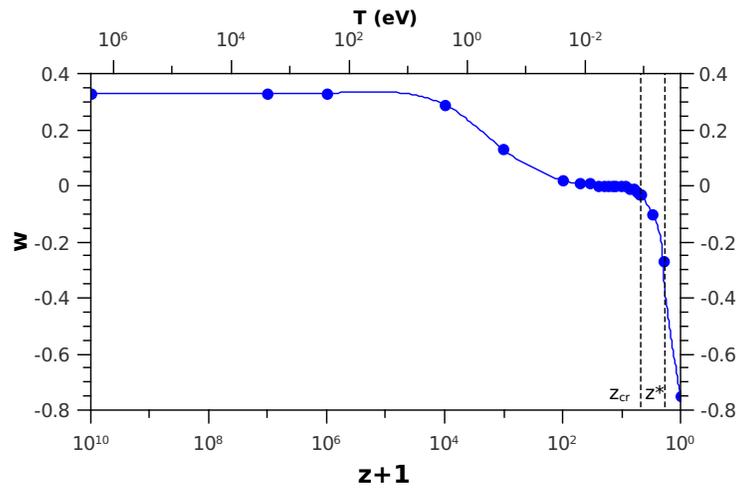,width=0.60\textwidth} \caption{(Color online)
Equation of state parameter
$w_{\mathrm{tot}}= P_{\mathrm{tot}}/  \rho_{\mathrm{tot}}$
for $M=2.39 \cdot 10^{-3}$~eV ($\alpha=0.01$) plotted up to the current redshift
(temperature, upper axis). }\label{wtot}
\end{figure}

The equation of state parameter of the entire Universe, $w_{\mathrm{tot}}$, is given by
$P_{\mathrm{tot}}= w_{\mathrm{tot}} \rho_{\mathrm{tot}}$. Since the matter contribution
$P_{M}=0$, then $P_{\mathrm{tot}}=P_\gamma + P_{\varphi \nu}$, where
$P_\gamma = \frac13 \rho_\gamma$ and the pressure of the coupled model
$P_{\varphi \nu}$ is obtained from Eq.~(\ref{OmR}). The numerical results of
$w_{\mathrm{tot}}$ are given in Fig.~\ref{wtot}.
%
%
%

To analyze the dynamics of the coupled model we need, in principle, to go
beyond the saddle-point approximation applied in the previous sections and
solve the equation of motion:
\begin{equation}
\label{EqMotion}
    \ddot{\varphi}+ 3 H \dot{\varphi}+\frac{\partial \Omega}{\partial
    \varphi}=0~.
\end{equation}
Above the transition point ($T> T_{\mathrm{crit}}$) the dynamics is quite
simple. Let us analyze perturbations to the saddle-point solution of
(\ref{SSB}):
\begin{equation}
\label{psi}
    \varphi(t) \equiv \phi_c + \psi(t)~.
\end{equation}
Taylor-expanding the thermodynamic potential of the coupled model
\begin{equation}
\label{dOmegaExp}
    \frac{\partial \Omega}{\partial \varphi}= \omega^2 \psi+ \frac12
    \Omega'''(\phi_c) \psi^2+~...
\end{equation}
with $\omega^2 \equiv \Omega''(\phi_c)$, we obtain from (\ref{EqMotion}) the
equation of a damped harmonic oscillator to the leading order:
\begin{equation}
\label{Osc}
    \ddot{\psi}+ 3 H \dot{\psi}+\omega^2 \psi =0~.
\end{equation}
So, the quintessence field $\varphi(t)$ oscillates around its saddle-point
value $\phi_c$ with $\psi(t) \propto e^{\imath \omega t- \frac32 H t}$. The
damping is very small, since as one can check
\begin{equation}
\label{Damp}
    \omega \gg \frac32 H~.
\end{equation}
The violation of the above condition and  breaking down of the oscillating
regime occurs in the vicinity of the critical point,  which is the inflection
point of the potential ($\omega=0$). This is the well-known phenomenon of the
critical slowing down near phase transition. Retaining the first non-vanishing
term in (\ref{dOmegaExp}), the equation of motion in the vicinity of the
critical point reads:
\begin{equation}
\label{NearTc}
    \ddot{\psi}+ 3 H \dot{\psi}+\frac12 \Omega'''(\phi_c) \psi^2=0~.
\end{equation}
Neglecting the small damping term in this equation, its solution can be found
analytically via a hypergeometric function. Since the explicit form of this
solution is not very interesting at this point, we just emphasize the
qualitative conclusion of the analysis: the fluctuation $\psi(t)$ oscillates
near the classical field $\phi_c$ in the stable (metastable) phase at
$T>T_{\mathrm{crit}}$, and it enters the run-away (power-law) regime when $T
\to T^+_{\mathrm{crit}}$.\cite{Bautin}
%
%
\subsection{Late-Time Acceleration of the Universe. Towards the End of Times}\label{Now}
%
%
%
%
The equilibrium methods are not applicable below the phase transition, and we
study the dynamics of the model from the equation of motion (\ref{EqMotion})
together with the Friedmann equations
(\ref{friedmann1},\ref{friedmann2},\ref{Frid0}). Solution of the Dirac
equations yields  $\rho_s \propto a^{-3}$ for the chiral density
\cite{Micheletti:2010cm}, so the equation of motion (\ref{EqMotion}) at $a \leq
a_{\mathrm{crit}}$ reads:
\begin{equation}
\label{EqMotion2}
    \ddot{\varphi}+ 3 H \dot{\varphi}= -\frac{\partial U}{\partial
    \varphi}-  \rho_{s,\mathrm{crit}}
    \Big( \frac{a_{\mathrm{crit}}}{a} \Big)^3 ~.
\end{equation}
From the results of the previous section we evaluate the chiral density  at the
critical point:
\begin{equation}
\label{rhoc}
    \rho_{s, \mathrm{crit}} \approx \alpha \Big(\frac{\Delta_{\mathrm{crit}}}{\nu}
    \Big)^{\alpha+1}  M^3~.
\end{equation}
The system of the integro-differential equations
(\ref{EqMotion2},\ref{friedmann1},\ref{friedmann2},\ref{Frid0}) was solved
numerically. All the quantities entering those equations are defined in the
previous subsection, except that one needs to include the extra term $\frac 12
\dot{\varphi}^2$ in the computation of both $\rho_{\mathrm{tot}}$ and $P_{\mathrm{tot}}$.
However the numerical results show that in the regimes of the parameters we are
interested, the kinetic term can be safely neglected. Since the critical point
of the model lies in the matter-dominated regime (cf. Fig.~\ref{Omega}), we
start with the Hubble parameter $H=2/3t$ ($a \propto t^{2/3}$). At the latest
times ($z \lesssim 1$) the Hubble parameter was determined self-consistently
from the numerical solution of the Friedmann equations.

\begin{figure}[h]
\epsfig{file=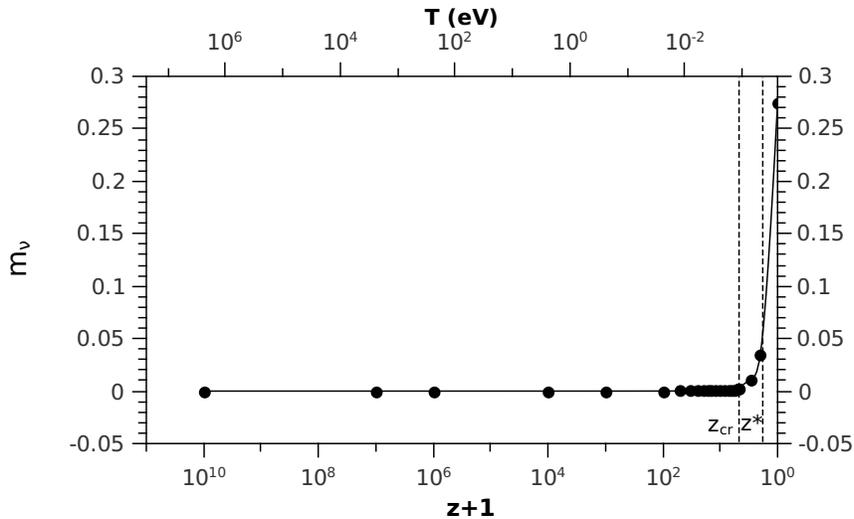,width=0.70\textwidth} \caption{Neutrino mass $m$
for $M=2.39 \cdot 10^{-3}$~eV ($\alpha=0.01$) plotted up to the current redshift
(temperature, upper axis). }\label{mnutotal}
\end{figure}

We find numerically that the quintessence field $\varphi(t)$ from the critical
point to the present time oscillates quickly (with the period $\tau \sim
10^{-27}$ Gyr) around the smooth (``mean value'') solution $\bar \varphi (t)$,
where the ``mean'' $\bar \varphi$ nullifies the r.h.s. of the equation of motion
(\ref{EqMotion2}). Relating the mean values with the physically relevant
observable quantities, we can easily obtain the key results analytically. (They
are checked against direct numerical calculations and found to be accurate
within 5 \% at most). Thus we get
\begin{eqnarray}
\label{fMD}
  &~& \bar \varphi = \varphi_{\mathrm{crit}} \cdot
     \Big( \frac{1+ z_{\mathrm{crit}}}{1+z}\Big)^{\frac{3}{\alpha+1}} ~, \\
\label{roMD}
 &~& \rho_{\bar \varphi}= \rho_{\varphi,\mathrm{crit}} \cdot
     \Big( \frac{1+ z}{1+z_{\mathrm{crit}}}\Big)^{\frac{3 \alpha}{\alpha+1}} ~,
\end{eqnarray}
where $\varphi_{\mathrm{crit}} \approx \frac{\nu}{\Delta_{\mathrm{crit}}} M$
and $\rho{_{\varphi,\mathrm{crit}}} \approx
(\frac{\Delta_{\mathrm{crit}}}{\nu})^\alpha M^4$. Having a free model parameter
$M$, we'll set it by matching the current density of the
scalar field $\rho_{\varphi, \mathrm{now}}$ to the observable value of the DE
density (\ref{DEnow}), so
\begin{equation}
\label{MDM}
    M= \big(\nu^\alpha \rho_{\varphi,\mathrm{now}} \big)^{\frac{\alpha+1}{\alpha+4}}
        \Delta_{\mathrm{crit}}^{-\alpha}
        T_{\mathrm{now}}^{-\frac{3\alpha}{\alpha+4}} ~.
\end{equation}
The exponent of the quintessence potential $\alpha $ is now the only parameter which can
be varied. We define the time-dependent mass via the solution of the motion equation as
$m(t)=\bar \varphi(t)$, thus obtaining an estimate for the present-time neutrino mass.
Results for various $\alpha$ are given in Table~\ref{Parameters}. There we also calculate
the critical points parameterized by the redshifts $z_{\mathrm{crit}}$ and the crossover
points $z^*$. The latter is defined as the redshift at which the Universe starts its
late-time acceleration, i.e., where $w_{\mathrm{tot}}=-\frac13$.
%
%
\begin{table}[h]
  \caption{Model's parameters and observables for various  $\alpha$.
  All the entries in this table are defined in the text.}
  ~\\
\begin{tabular}{|c|c|c|c|c|}
  \hline
  ~ & ~ & ~ & ~ & ~ \\[-0.4cm]
  $~~\alpha~~$ & $~~M$ (eV)~~ & $~m_{\mathrm{now}}$ (eV)~ & $~~z_{\mathrm{crit}}~~$ & $~~z^*~~$ \\
  ~ & ~ & ~ & ~ & ~ \\[-0.4cm]
  \hline
  2 &     $~9.75 \cdot 10^{-2}~$ & 167 & 392 & 4.9 \\
  \hline
  1 &     $~1.69 \cdot 10^{-2}~$ & 44.6 & 76.6 & 2.3 \\
  \hline
  1/2 &   $~6.33 \cdot 10^{-3}~$ & 17.0 & 27.7 & 1.5 \\
  \hline
  $10^{-1}$ &  $~2.81 \cdot 10^{-3}~$ & 2.82 & 8.73 & 0.93 \\
  \hline
  $~10^{-2}~$  & $~2.39 \cdot 10^{-3}~$ & 0.27 & 3.67 & ~0.83~ \\
  \hline
  $~10^{-3}~$  & $~2.36 \cdot 10^{-3}~$ & 0.027 & 1.60 & ~0.82~ \\
  \hline
\end{tabular}
 \label{Parameters}
\end{table}
%
%
For the present time we find
\begin{equation}
 \label{wtotnow}
    w_{\mathrm{tot}}^{\mathrm{now}} \approx -\frac34~.
\end{equation}
As we infer from the data of Table~\ref{Parameters}, the range of exponents
$\alpha \ll 1$ corresponds to more realistic predictions for the neutrino
mass \cite{Dolgov,Mohapatra07,Avignone:2007fu} and for the crossover redshift
$z^*$  \cite{Ishida:2007ej}.
For $\alpha =0.01$ we plot the evolution of the relative energy densities, the
equation of state parameter, and the neutrino mass in
Figs.~\ref{Omega},\ref{wtot},\ref{mnutotal}.
%
%
%
%

We consider the quite artificial case of small quintessence exponent $\alpha$
as an ansatz crossing over smoothly from physically plausible potentials with,
say, $\alpha = 1$ or 2 to the logarithmic potential
\begin{equation}
\label{LogU}
    U(\varphi)=M^4 \Big( 1+ \alpha \log \frac{M}{\varphi} \Big)~.
\end{equation}
The latter often appears in various contexts
\cite{Copeland:2006wr,Fardon:2003eh}.\footnote{The numerical results for small
parameter $\alpha$, as e.g. $\alpha =0.01$ taken for the plots, are virtually
indistinguishable for the cases of the Ratra-Peebles (\ref{RP}) or logarithmic
(\ref{LogU}) potentials. However the Ratra-Peebles potential at more
``natural'' $\alpha=1,2$ allows to probe the coupled fermionic-quintessence
models in the search of heavy DM particle candidates. }

%
%
%
\section{Conclusions}\label{Concl}
%
%
%
In this paper we analyzed the MaVaN scenario in a framework of a simple
``minimal'' model with only one species of the (initially) massless Dirac
fermions coupled to the scalar quintessence field. By using the methods of
thermal quantum field theory we derived for the first time (in the context of
the MaVaN or, even more broadly, the VAMP models) a consistent equation for
fermionic mass generation in the coupled model.

We demonstrated that the mass equation has non-trivial solutions only for special
classes of potentials and only within certain temperature intervals.
It appears that these results have not been reported in the literature on
VAMPs before now.

We gave most of the results for the particular choice of a trial DE potential
-- the Ratra-Peebles quintessence potential. This potential has all the necessary
properties we needed for our task: it is simple, it satisfies the criteria we
found for non-trivial solutions of the mass equation to exist, and it has only
one dimensionfull parameter- the energy scale $M$ to tune. Also, at small values
of the exponent $\alpha$ it effectively crosses over to the case of a logarithmic potential.
We have checked that other potentials, e.g., exponential, lead to a qualitatively
similar picture, but they have at least one more energy scale to handle, which we
consider as an unnecessary complication at this point.

We analyzed the thermal (i.e. temporal) evolution of the model, following the time
arrow. Contrary to what one might expect from analogies with other contexts, like,
e.g., condensed matter, the model does not generate the mass via a conventional
spontaneous symmetry breaking below a certain temperature. Instead it has a non-trivial
solution for the fermionic  mass evolving ``smoothly''  from zero at the ``point''
$T=\infty$. The scalar field is infinitely heavy at the same point.
More realistically, we assumed the model is applicable starting at the temperatures
somewhere in the beginning of the radiation-dominated era. We found that the DE
contribution in this regime is subleading, and the model behaves as an ultra-relativistic
Fermi gas at those temperatures.

This regime corresponds to a stable phase of the model given by a global minimum of
the thermodynamic potential $\Omega(\varphi)$.  The temperature/time dependent minimum
$\langle  \varphi \rangle$ generates the varying fermionic mass
$m \propto \langle  \varphi \rangle$.

With increase in time, as the temperature decreases, the model reaches the point of
metastability where its pressure ($P$) vanishes. From our estimates of the model's
scales, we showed that this happens during the matter-dominated era of the Universe.
At this point the system's ground state becomes doubly degenerate, and the potential
$\Omega=0$ at the non-trivial (finite) minimum  $\langle  \varphi \rangle$
as well as at the trivial vacuum $ \varphi= \infty$.

Further on, at lower temperatures the system stays in the metastable (supercooled)
state until it reaches the critical point where the local minimum of the
thermodynamic potential disappears and it becomes an inflexion point. At this critical
temperature the model undergoes a first-order (discontinuous) phase transition.
At the critical point \textit{the equilibrium values} of the fermionic and the
scalar field masses discontinuously jump to the `doomsday'' vacuum  state values
$m= \infty$ and $m_\phi=0$, respectively. The square of the sound velocity and
equation of state parameter $w$ have \textit{the equilibrium values} corresponding
to the de Sitter Universe with a cosmological constant, i.e. $c^2_s=w=-1$. It is worth
pointing out that $c^2_s>0$ in both the stable and metastable phases, and the
sound velocity vanishes reaching the critical temperature from above.

Since the equilibrium approach is not applicable below the critical temperature,
we find parameters of the model from  direct numerical solution of the equation of motion
and the Friedmann equations. The single scale $M$ of the quintessence potential
is chosen to match the present DE density, then other parameters of the Universe
are determined. We obtain a consistent picture: the phase transition has
occurred rather recently at $z_{\mathrm{crit}} \lesssim 5$  during the
matter-dominated era, and the Universe is now being driven towards the stable vacuum
with zero $\Lambda$-term. The expansion of the Universe accelerates
starting from $z^* \approx 0.83$. Setting  $\alpha =0.01$ for
$M \approx 2.4 \cdot 10^{-3}~\mathrm{eV}$, we end up with the neutrino mass
$m \approx 0.27~\mathrm{eV}$.

The present results allow us to propose a completely new viewpoint not only on
the MaVaN, but on the quintessence scenario for the Universe as well. The
common concerns about the slow-rolling mechanism for the DE relaxation toward
the $\Lambda=0$ vacuum are related to the question of what is the mechanism to
set the initial value of the scalar field $\varphi$ where it evolves (rolls
down) from. Our results demonstrate that up to recent times (i.e. above the
critical temperature) the quintessence field was locked around its average
(classical) value $\langle  \varphi \rangle$. Its value is determined by the
scale $M$ and the temperature. The average $\langle  \varphi \rangle$ gives the
fermionic mass at the same time. The scalar field is rigid (i.e. massive),
although it softens (i.e., its mass decreases) as the system approaches the
critical temperature. Above the critical temperature the scalar field can only
oscillate around its equilibrium value $\langle \varphi \rangle$. At the
critical point the minimum of the thermodynamic potential becomes the inflexion
point, the scalar field looses its rigidity (mass). Then the field can only
roll down towards the new stable ground state $\Omega=0$ at $\varphi=\infty$.
So physically, the critical point corresponds to the transition of the Universe
from the stable oscillatory to the unstable rolling regime.

A more sophisticated numerical study of the kinetics after the critical
point is warranted in order to address such issues as the detailed description
of the crossover between different regimes, and the clustering of neutrinos.
These and some other questions are relegated to our future work.

%
%
%
\begin{acknowledgments}
We highly appreciate useful comments and discussions with D. Marfatia, B.
Ratra, and N. Weiner. We are grateful to N. Arhipova, D. Boyanovsky, R. Brandenberger, O.
Chkvorets, H. Feldman, A. Gruzinov, L. Kisslinger, the late L. Kofman, S.
Lukyanov, and U. Wichoski for helpful discussions and communications. We thank
the anonymous referee for constructive criticism and comments which stimulated
us to undertake deeper analyses of the model, and especially of its dynamics.
G.Y.C. thanks the Center for Cosmology and Particle Physics at New York
University for hospitality. We acknowledge financial support from the Natural
Science and Engineering Research Council of Canada (NSERC), the Laurentian
University Research Fund (LURF), Scientific Co-operation Programme between
Eastern Europe and Switzerland (SCOPES), the Georgian National Science
Foundation grants \# ST08/4-422.  T.K. acknowledges the support from NASA
Astrophysics Theory Program grant NNXlOAC85G and the ICTP associate membership
program. A.N. thanks the Bruce and Astrid McWilliams Center for Cosmology for financial support.
\end{acknowledgments}
%


%
\end{document}